\documentclass[5p, times]{elsarticle}
\usepackage[utf8]{inputenc}

\usepackage{amssymb}
\usepackage{amsthm}
\usepackage{amsmath}
\usepackage{lineno}
\usepackage{xcolor}
\usepackage{enumitem}
\usepackage{textcomp}

\newcommand{\xpol}[1]{{\sf XPOL#1}}
\newcommand{\xpoli}{\xpol{-I}}
\newcommand{\xpoliii}{\xpol{-III}}

\begin{document}

\begin{frontmatter}

\title{\xpoliii: a New-Generation VLSI CMOS ASIC for High-Throughput X-ray Polarimetry}

\author[1]{Minuti M.}
\author[2,1]{Baldini L.}
\author[1]{Bellazzini R.}
\author[1]{Brez A.}
\author[1]{Ceccanti M.}
\author[3]{Krummenacher F.}
\author[4]{Latronico L.}
\author[1]{Lucchesi L.}
\author[1]{Manfreda A.}
\author[1]{Orsini L.}
\author[1]{Pinchera M.}
\author[1]{Profeti A.}
\author[1]{Sgrò C.}
\author[1]{Spandre G.}
\address[1]{Istituto Nazionale di Fisica Nucleare, Sezione di Pisa, Largo B. Pontecorvo 3, I-56127 Pisa, Italy}
\address[2]{Università di Pisa, Dipartimento di Fisica Enrico Fermi, Largo B. Pontecorvo 3, I-56127 Pisa, Italy}
\address[3]{Advanced Silicon SA, Lausanne, Switzerland}
\address[4]{Istituto Nazionale di Fisica Nucleare, Sezione di Torino, Via P. Giuria, 1, I-10125 Torino, Italy}

\journal{NIM A}
\date{Compiled on \today}

%\tnotetext[t1]{This work was supported by the Italian Space Agency (ASI) through the agreements ASI-INAF-2018-11-HH.O and ASI-INFN-2017.13-H0.}

\begin{abstract}
While the successful launch and operation in space of the Gas Pixel Detectors onboard the 
PolarLight cubesat and the Imaging X-ray Polarimetry Explorer demonstrate the viability
and the technical soundness of this class of detectors for astronomical X-ray polarimetry, 
it is  clear that the current state of the art is not ready to meet the challenges of the 
next generation of experiments, such as the enhanced X-ray Timing and Polarimetry mission,
designed to allow for a significantly larger data throughput.

In this paper we describe the design and test of a new custom, self-triggering readout ASIC,
dubbed \xpoliii, specifically conceived to address and overcome these limitations. While building 
upon the overall architecture of the previous generations, the new chip improves over its 
predecessors in several, different key areas: the sensitivity of the trigger electronics,
the flexibility in the definition of the readout window, as well as the maximum speed for
the serial event readout. These design improvements, when combined, allow for almost an order
of magnitude smaller dead time per event with no measurable degradation of the polarimetric,
spectral, imaging or timing capability of the detector, providing a good match for the next 
generation of X-ray missions.
\end{abstract}

\begin{keyword}
X-ray polarimetry
\PACS 95.55.Ka \sep 95.55.Qf
\end{keyword}

\end{frontmatter}

\section{Introduction}
\label{sec:introduction}

The recent launches of the PolarLight cubesat~\cite{PolarLight} and the Imaging X-ray
Polarimetry Explorer (IXPE)~\cite{IXPEJATIS} signal the re-opening of a new observational 
window---that of astronomical X-ray polarimetry---after almost 40 years. At the heart of
both missions are innovative polarization-sensitive Gas Pixel Detectors (GPD)~\cite{BALDINI2021102628},
exploiting the photoelectric effect to derive the linear polarization of the
incoming radiation based on the reconstruction of the azimuthal direction of emission of 
the photo-electrons. The design, qualification and successful operation in space of the
PolarLight and IXPE Gas Pixel Detectors mark the culmination of a long R\&D 
program~\cite{Costa2001662, BELLAZZINI2004477, BELLAZZINI2006552}, following the first 
pioneering attempts at a practical implementation of photoelectric X-ray polarimetry 
(see, e.g., \cite{Ramsey}).

Future X-ray missions will pose even tighter requirements on focal-plane detectors. 
With a much larger effective area (about a factor $\times~5$ compared to IXPE) the mirror 
modules of the Polarimetry Focusing Array (PFA) onboard the enhanced X-ray Timing and 
Polarimetry (e-XTP) mission~\cite{eXTP} will produce too large of a data throughput for
the current generation of Gas Pixel Detectors, even for moderately bright sources.
It is then clear that 
%, in any realistic scenario, 
a drastic reduction of the dead-time per event
is one of the necessary preconditions for a significant leap in sensitivity for the next
generation of polarimetric missions.

In this paper we describe \xpoliii, a new-generation custom CMOS ASIC specifically 
designed for high-throughput X-ray polarimetry applications, taking full advantage of the 
lessons learned through the development of the IXPE mission. The new design is specifically
aimed at a substantial reduction of the average dead time per event, while preserving all 
the other relevant high-level performance metrics of the readout chip. As detailed in
section~\ref{sec:performance}, the first comprehensive test campaign that we performed
on the new ASIC confirms that all design goals were met, and \xpoliii{} is a prime candidate
to provide polarimetric capabilities to future X-ray missions.

\section{\xpoliii\ Architecture}
\label{sec:architecture}

The main features of the previous \xpol\ generations, most of which are also relevant 
in the context of this work, are thoroughly described 
in~\cite{BELLAZZINI2004477, BELLAZZINI2006552} and~\cite{BALDINI2021102628}.
In this section we only provide a succinct summary of the internal functioning of the chip, 
with emphasis on the design changes specific to \xpoliii.
%in order to set the stage for the discussion of the test results that we shall present in the following.

\begin{figure}[ht!]
    \centering
    \includegraphics[width=0.75\linewidth]{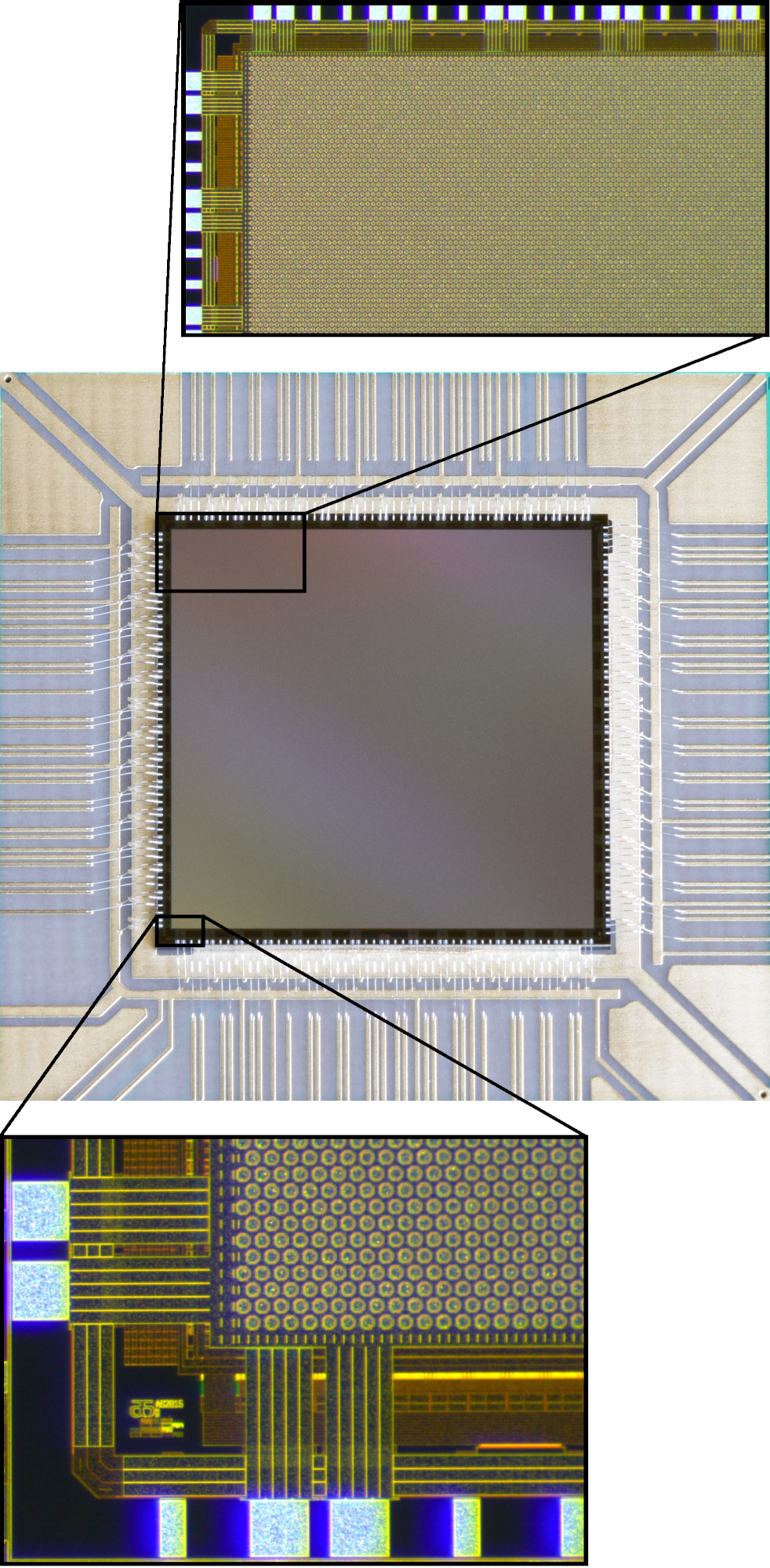}
    \caption{Picture of a \xpoliii\ chip (center panel), glued and wire-bonded onto its 
    printed circuit board. The insets show micro-photographs of two of the corners at 
    different magnification levels, with the triangular structure of the pixel array clearly visible 
    in the bottom panel.}
    \label{fig:xpol3}
\end{figure}

\begin{table}[htb!]
    \centering
    \begin{tabular}{p{0.5\linewidth}p{0.45\linewidth}}
    \hline
    Parameter &	Value\\
    \hline
    \hline
    Number of pixels & $107\,008~(304 \times 352)$\\
    Physical pitch & 50~$\mu$m\\
    Shaping time & 1~$\mu$s\\
    Pixel gain & $200$~mV~fC$^{-1}$\\
    Pixel Noise & $30$~$e^{-}$~ENC\\
    Full scale linear range (FSLR) & $30\rm{k}~e^{-}$\\    
    Minimum trigger threshold & $\sim 150$~$e^{-}$ (0.5\% of FSLR)\\
    \hline
    \end{tabular}
    \caption{Summary table of the basic geometrical and electrical characteristics
    of the \xpoliii{} readout ASIC.}
    \label{tab:asic_characteristics}
\end{table}

\begin{figure}[bht!]
    \centering
    \includegraphics[width=\linewidth]{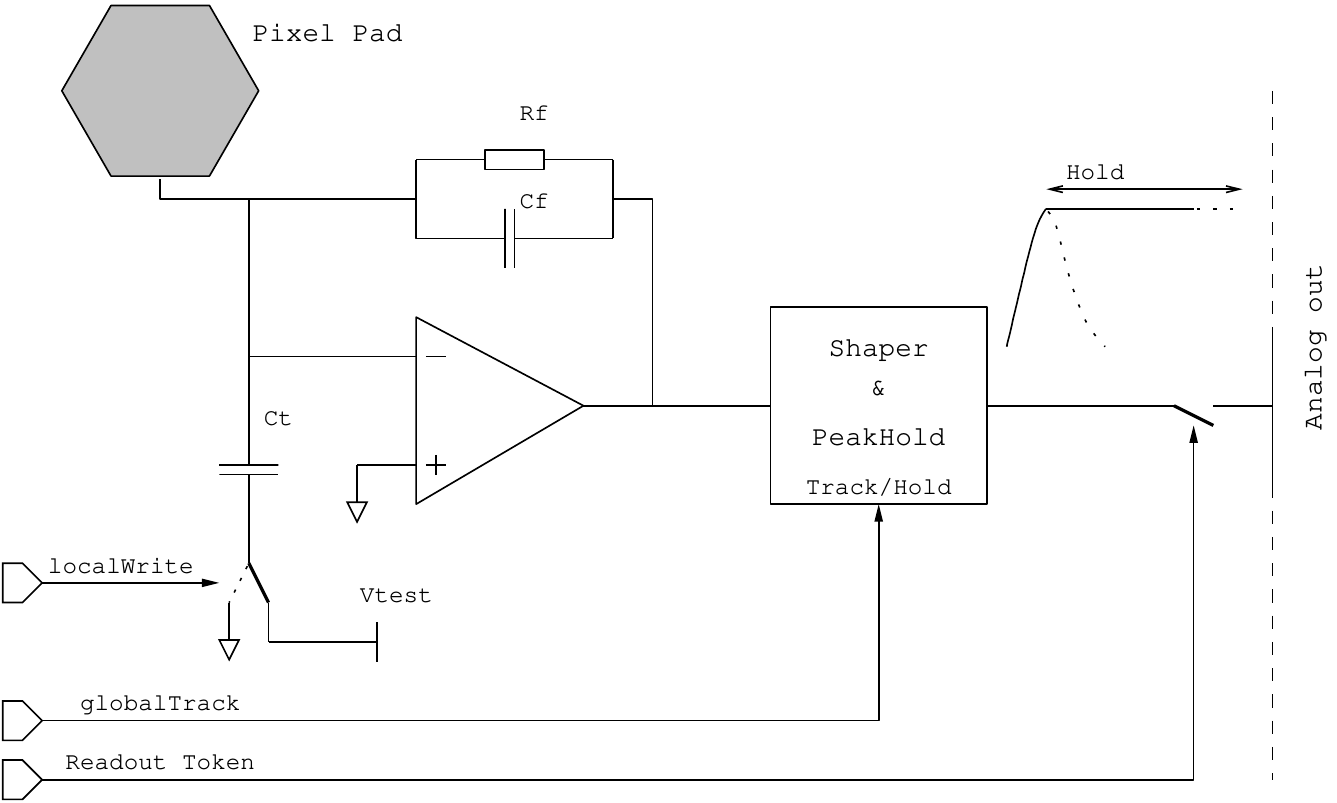}
    \caption{Simplified pixel schematic.}
    \label{fig:pixel_schematic}
\end{figure}

Manufactured with a standard 180~nm CMOS process, the \xpoliii\ readout ASIC (shown in 
Figure~\ref{fig:xpol3}) is organized as a rectangular  matrix of $107\,008$~hexagonal pixels
arranged in a tringular pattern of~$304$ columns and $352$ rows at $50$~$\mu$m pitch, for a total 
active area of $15.2 \times 15.2$~mm$^2$. Each pixel is composed of a metal electrode (from the 
top layer of the process), acting as a charge-collecting anode, connected to a charge-sensitive
amplifier, followed by a shaping circuit and a sample-and-hold system, as illustrated in the 
simplified schematic in Figure~\ref{fig:pixel_schematic}.

Similarly to the previous ASIC generations, \xpoliii{} provides an advanced self-triggering
capability, with automatic localization of the \emph{region of interest} (ROI) containing the 
photo-electron track. Upon trigger, the outputs of the pixels within the ROI are sequentially
connected to an on-chip global differential amplifier driving the output pads, and the digitization of the corresponding signal is performed with a dedicated, external ADC. 
At a very fundamental level, this is the basic mechanism that allows to keep the overall 
readout time manageable, as, for a typical event, we only read out a few hundreds pixels out
of the $> 100$~k in the full matrix.
Building on this concept, in the initial stages of the \xpoliii\ design we identified three distinct possible
lines of actions to speed up the readout:
\begin{itemize}[leftmargin=10pt]
    \item reduce the average size of the region of interest;
    \item increase the maximum frequency of the serial readout clock;
    \item streamline the readout sequence to avoid unnecessary delays.
\end{itemize}
As we shall discuss in the remaining of this section, the combination of these three, 
seemingly simple updates provides a drastic dead-time reduction.

\subsection{Event Triggering}
\label{sec:trigger}

Every 4 pixels in the ASIC are logically OR-ed together to contribute to a local trigger
with a dedicated shaping amplifier. This basic building block of $2 \times 2$ pixels is 
called a trigger \emph{mini-cluster}, and is central to the entire trigger logic.
Upon trigger, the event is automatically localized by the ASIC core logic in the smallest
rectangle containing all triggered mini-clusters, called the \emph{region of trigger} (ROT).
More specifically, the chip calculates and makes available in a specific register the coordinates
$\left<X_\text{min},~Y_\text{min}\right>$ and  $\left<X_\text{max},~Y_\text{max}\right>$ of the
upper-left and lower-right corners of the ROT. On an event by event basis, this information 
can be manipulated in an arbitrary fashion---typically by adding a pre-defined padding on the
four borders---to define the \emph{region of interest} (ROI) for the event capture and readout. 

\begin{figure*}[ht!]
    \centering
    \includegraphics[width=0.8\textwidth, clip=true, trim=50 90 0 20]{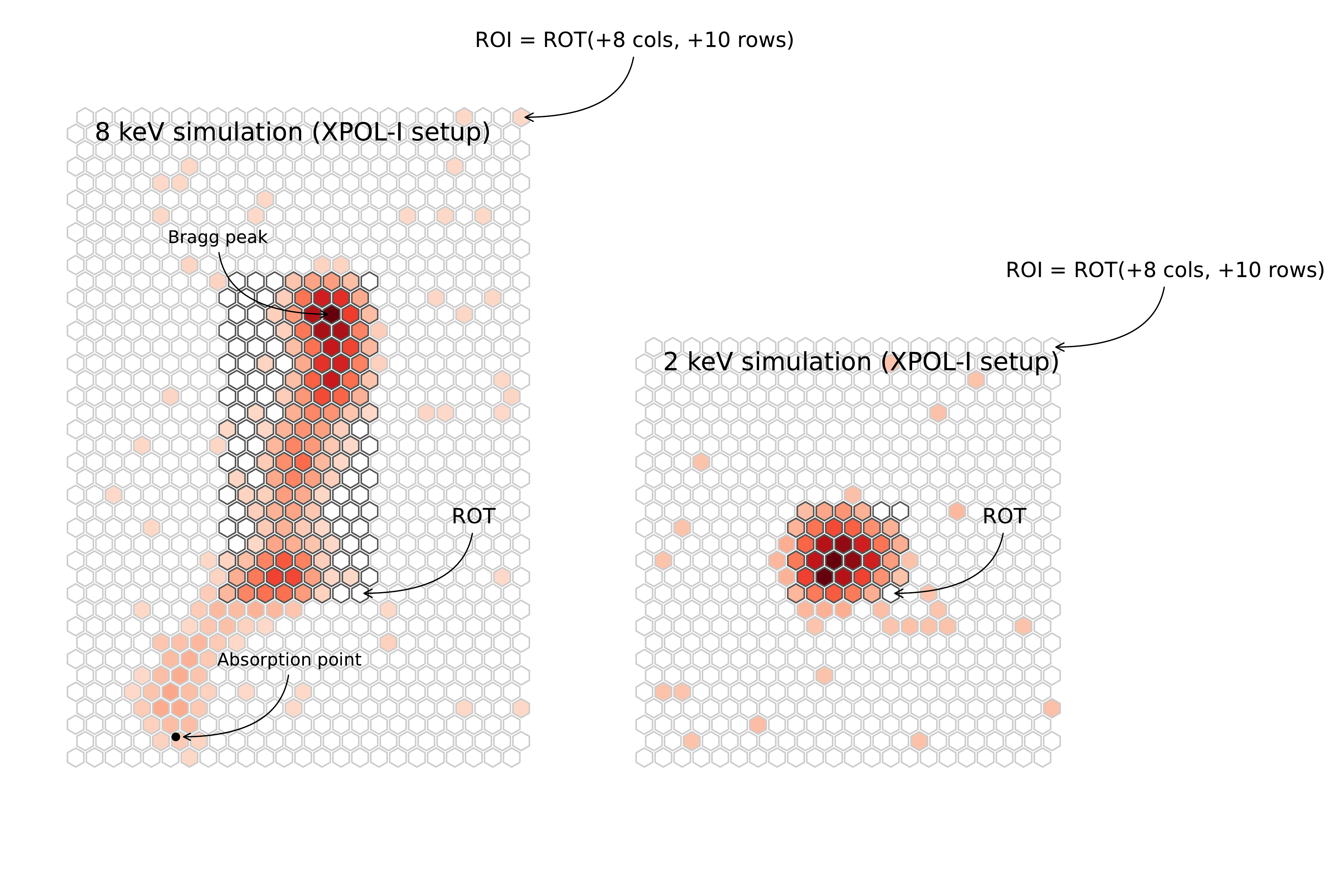}
    \caption{Sample simulated photo-electron tracks at 8~keV (left) and 2~keV (right)
    in pure DME at 800~mbar, embedded in the respective \xpoli-like readout windows at 
    the nominal trigger threshold. Pixels are depicted with black and gray borders depending
    on whether they lie within the ROT or the outer padding region (8 columns on the left and
    the right of the ROT and 10 rows on the top and the bottom).
    As illustrated in the left panel, this comparatively large padding is necessary in the
    \xpoli\ setup in order to guarantee that the initial part of high-energy tracks is fully
    contained in the ROI, but is largely sub-optimal on the low-end of the spectrum, where
    tracks are much more compact. With a significantly lower trigger threshold and a more
    flexible padding strategy, \xpoliii\ is capable of generating much smaller ROIs
    without compromising the track containment (see~Figure~\ref{fig:xpoliii_track}).}
    \label{fig:xpoli_tracks}
\end{figure*}

While the top-level trigger logic in the previous generations of the ASIC was similar to that
of \xpoliii, the algorithm implemented in \xpoli\ to define the ROI was comparatively crude:
a constant padding of 8~columns on the left and right and 10~rows on the top and bottom of the
ROT was automatically added by the ASIC, as shown in Figure~\ref{fig:xpoli_tracks}, with no 
room for additional tuning. This approach served well the primary purpose of making the
ROI big enough to fully contain the tracks at all the energies of interest (and was ultimately adequate for the typical IXPE throughput), but was clearly sub-optimal at low energy, where
tracks tend to be more compact and with a comparatively higher ionization density.

In contrast, \xpoliii\ offers maximal flexibility---to the point that, once the ROT has been
defined by the ASIC, the ROI can be calculated by means of any arbitrary external logic and
loaded into the proper register on an event-event-basis. Additionally, a \emph{hybrid} 
operational mode is available where a pre-loaded padding, independently adjustable on the 
four sides of the ROT, is automatically applied by the chip without the need of an additional
data transactions. (In practice, since the information available at the time that the ROI 
decision has to be made is limited, this hybrid mode is in fact outperforming overly 
sophisticated padding schemes, and will be the baseline configuration used for all the tests
described in the following of the paper.) Even more importantly, this additional flexibility in
the definition of the ROI is accompanied by another fundamental design change: the AC nature
of the \xpoliii{} trigger coupling. Compared to the \xpoli{} architecture, where the dispersion
of the DC offsets across different amplifiers was playing a prominent role, this change allows
to lower dramatically the minimum trigger threshold practically achievable, increasing, 
in turn, the fraction of mini-clusters in the track participating in the trigger. 

\begin{figure}
    \centering
    \includegraphics[width=\linewidth]{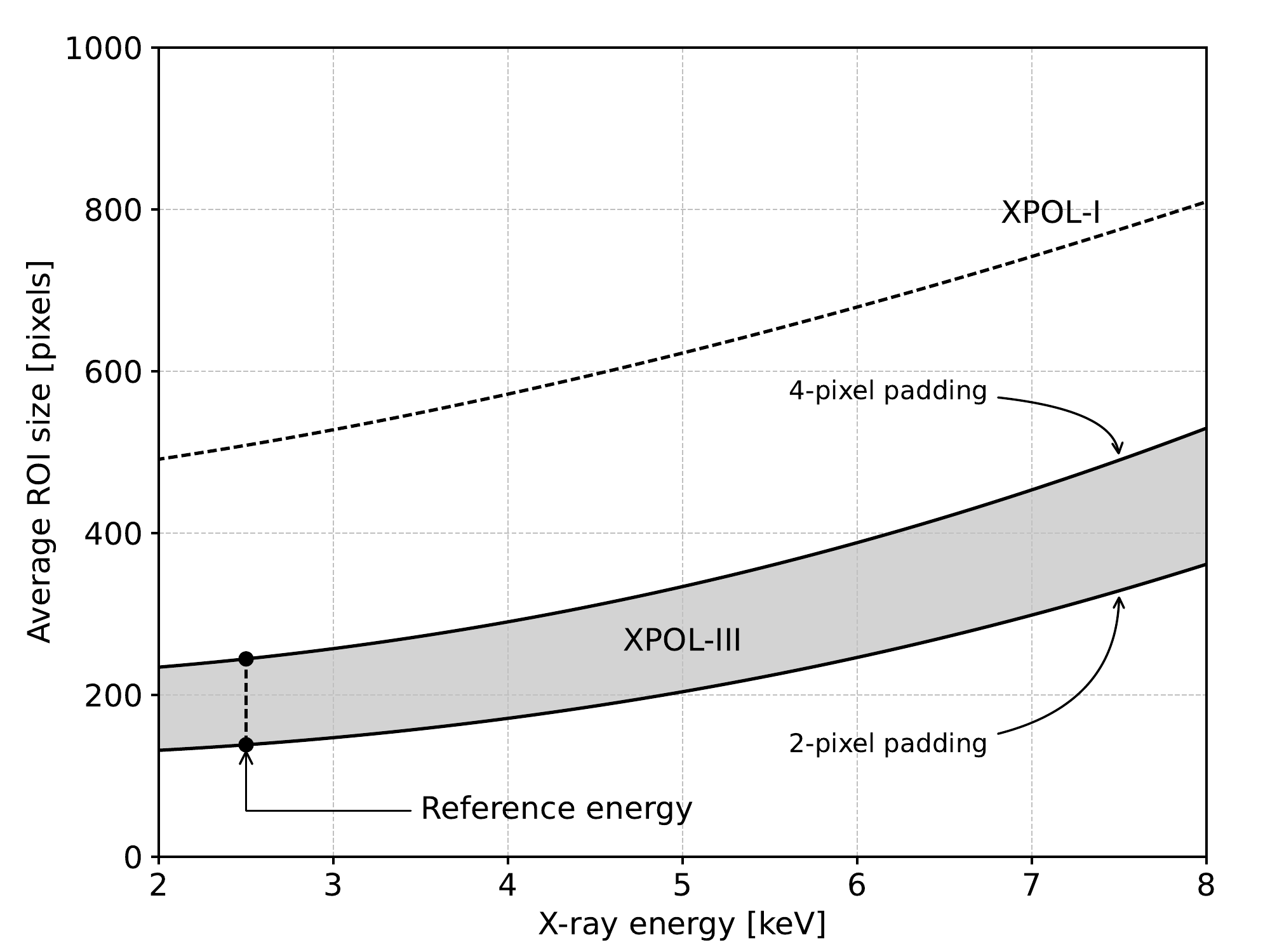}
    \caption{Average ROI size as a function of the X-ray energy for \xpoli{} and \xpoliii{}
    (in the latter case, with a constant padding of 2~and 4 pixels on all four sides of the
    ROT). As we shall see in section~\ref{sec:tests}, the shaded region between the two solid
    lines is the useful phase space for determining the optimal working point.
    For reference, at 2.5~keV (which is germane to the IXPE peak sensitivity), \xpoliii{} 
    yields an average ROI size comprised between 140 and 240 pixels, at the two different
    padding settings, which is 3.5--2 times smaller than the \xpoli{} equivalent.}
    \label{fig:roi_vs_energy}
\end{figure}

It cannot be over-emphasized that these two seemingly simple changes in the trigger circuitry
constitute a change of paradigm and a radical departure from how previous iterations of the 
ASIC operated. To put things in the simplest possible way, in \xpoli\ it was mostly the Bragg
peak to participate into the trigger, and we had to resort to a comparatively large padding 
of the ROT in order to capture the initial part of the high-energy tracks.
In \xpoliii\ the entire track participates into the trigger at all energies, and a minimal
additional padding is sufficient to ensure full containment of the track.
As shown in  Figure~\ref{fig:roi_vs_energy}, this allows for ROIs that are smaller on average,
and scale more favorably with the track length, offering additional leverage for the operation
at the focus of a X-ray optics, where the vast majority of the photons are detected at the 
lower end of the energy spectrum. Assuming a constant padding of 2 (4) pixels on all four sides
of the ROT, the \xpoliii\ design results in an overall reduction of the ROI size by a factor
of 3.5 (2), across the band. 

\subsection{Event Readout and Pedestal Subtraction}
\label{sec:readout}

Upon trigger, in nominal data-taking configuration, the chip autonomously initiates the 
peak-detection process of the internal sample-and-hold system, at the end of which the analog
output of each pixel within the ROI is sequentially routed to the differential output buffer
connected to the external ADC, and the serial readout proceeds driven by an adjustable readout
clock provided by the back-end electronics. The region of interest is actually read out twice,
and the FGPA on the DAQ board then performs the pixel-by-pixel pedestal subtraction and proceeds
to the zero suppression and the event compression and transmission~\cite{BarbaneraTNS}. As explained in~\cite{BALDINI2021102628}, this online pedestal subtraction is useful to minimize
subtle systematic effects that were found to be important for our application. 

While a detailed technical description of the readout sequence is beyond the scope of this 
paper, the readout time per event can be parametrized, neglecting sub-dominant contributions,
as a linear function of the number of pixels $n_{\mathrm pix}$ in the region of interest
\begin{align}\label{eq:readout_time}
    T_\mathrm{read} = q + m n_\mathrm{pix},
\end{align}
where the constant term $q$ incorporates, e.g., the peak-detection interval, the necessary 
data transactions to execute the readout and reset the system, as well as all the fixed 
delays that are needed to synchronize the sequence, while the slope $m$ captures all the
contributions whose duration is proportional to the size of the ROI, such as the event 
serial readout sequence and the pedestal subtraction in FPGA.
(We note that both $q$ and $m$ depend on the particular settings of the readout ASIC and the 
back-end electronics with, e.g., $m$ being dominated by the serial readout clock.)

Compared to \xpoli{}, much of the differential output has been redesigned in \xpoliii{} in 
order to streamline the readout. The design of the output buffer has been tuned to minimize 
the settling time of the analog signal and increase the maximum serial readout clock practically
usable, which for \xpoli{} was limited to $\sim 5$~MHz, and the analog reset circuitry has been
reworked to avoid the need for extra delays, necessary in \xpoli{} to avoid self-induced false
triggers.
One additional notable change is that the ROI coordinates, once determined, remain stored 
in their register until the latter is explicitly reset, which avoids another class of 
fixed delays necessary in \xpoli{} to perform the online pedestal subtraction.
The net result of all these optimization is a net (and significant) reduction of both terms 
in~Equation~\eqref{eq:readout_time}, as we shall see in section~\ref{sec:tests}.

\section{Measurement setup}
\label{sec:assembly}

\begin{figure}[htbp!]
    \centering
    \includegraphics[width=0.8\linewidth]{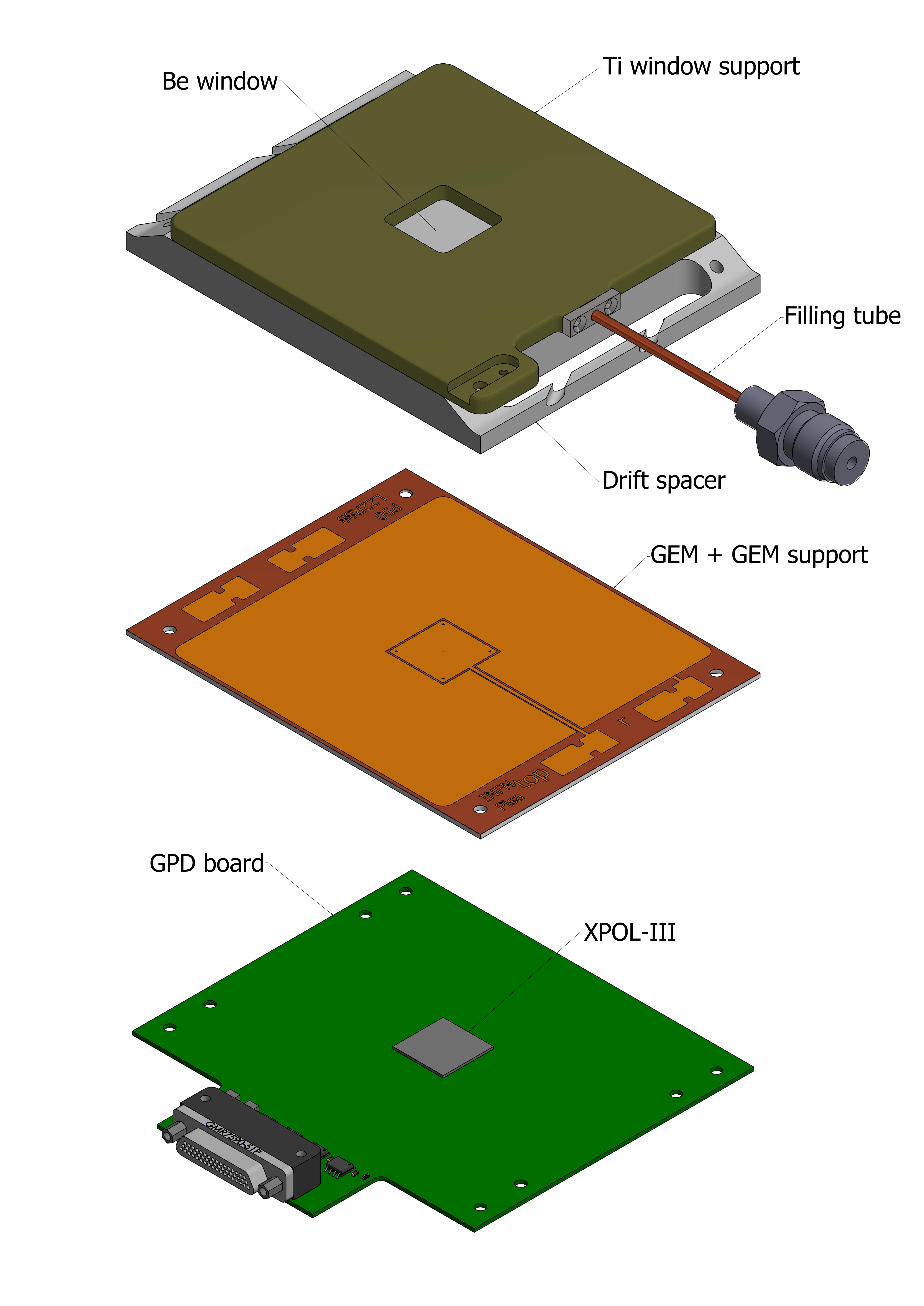}
    \caption{Exploded view of a GPD equipped with a \xpoliii{} readout ASIC. The main differences with respect to the models used in IXPE are the chip-on-board configuration and the absence of the internal alumination of the Be foil.}
    \label{fig:gpd_exploded}
\end{figure}

The GPD assembly is the natural \emph{packaging} for the readout ASIC, and an essential 
ingredient for testing its functionality beyond the basic electrical aliveness.
We have assembled, baked-out, filled and sealed entirely in house, a few detectors equipped with 
a \xpoliii{} readout ASIC.
Figure~\ref{fig:gpd_exploded} shows an exploded view of the detector, whose design implements some important simplification with respect to  the focal plane detectors onboard IXPE~\cite{BALDINI2021102628}.

We kept the top-level geometry and gas mixture, but we opted for a chip-on-board scheme 
(i.e. with the ASIC directly glued and wire-bonded over its printed circuit board, 
avoiding the use of a ceramic package) to simplify the assembly. 
We used $50~\mu$m pitch, $50~\mu$m thick GEM foils manufactured with a conventional chemical etching process. 
Finally, the entrance window is a simple $50~\mu$m Be foil with no alumination. 
The main characteristics of the detectors are summarized in table~\ref{tab:gpd_characteristics}.

\begin{table}[htb!]
    \centering
    \begin{tabular}{p{0.5\linewidth}p{0.4\linewidth}}
    \hline
    Parameter &	Value\\
    \hline
    \hline
    Entrance window & Pure Be, 50~$\mu$m\\
    Drift gap thickness & 10~mm\\
    Transfer gap thickness & 0.8~mm\\
    Readout configuration & Chip on board\\
    GEM pitch & 50~$\mu$m\\
    GEM hole diameter & 30~$\mu$m\\
    Gas filling & Pure DME @ 730~mbar\\
    \hline
    \end{tabular}
    \caption{Summary table of the basic properties of the GPD under test.}
    \label{tab:gpd_characteristics}
\end{table}

Each one of these modifications is a step toward our ultimate goal of a new-generation 
GPD, featuring a simpler and more robust mechanical assembly, and a higher 
polarization sensitivity---both in terms of modulation factor and uniformity 
of response to unpolarized radiation. A detailed description of the GPD development, 
including the description of a dedicated filling facility 
and the long-term stability of the performance, is beyond the scope of this work and 
will be  presented in a forthcoming paper.

\section{Electrical tests and working point definition}
\label{sec:tests}

\begin{figure}[!htb]
    \centering
    \includegraphics[width=\linewidth]{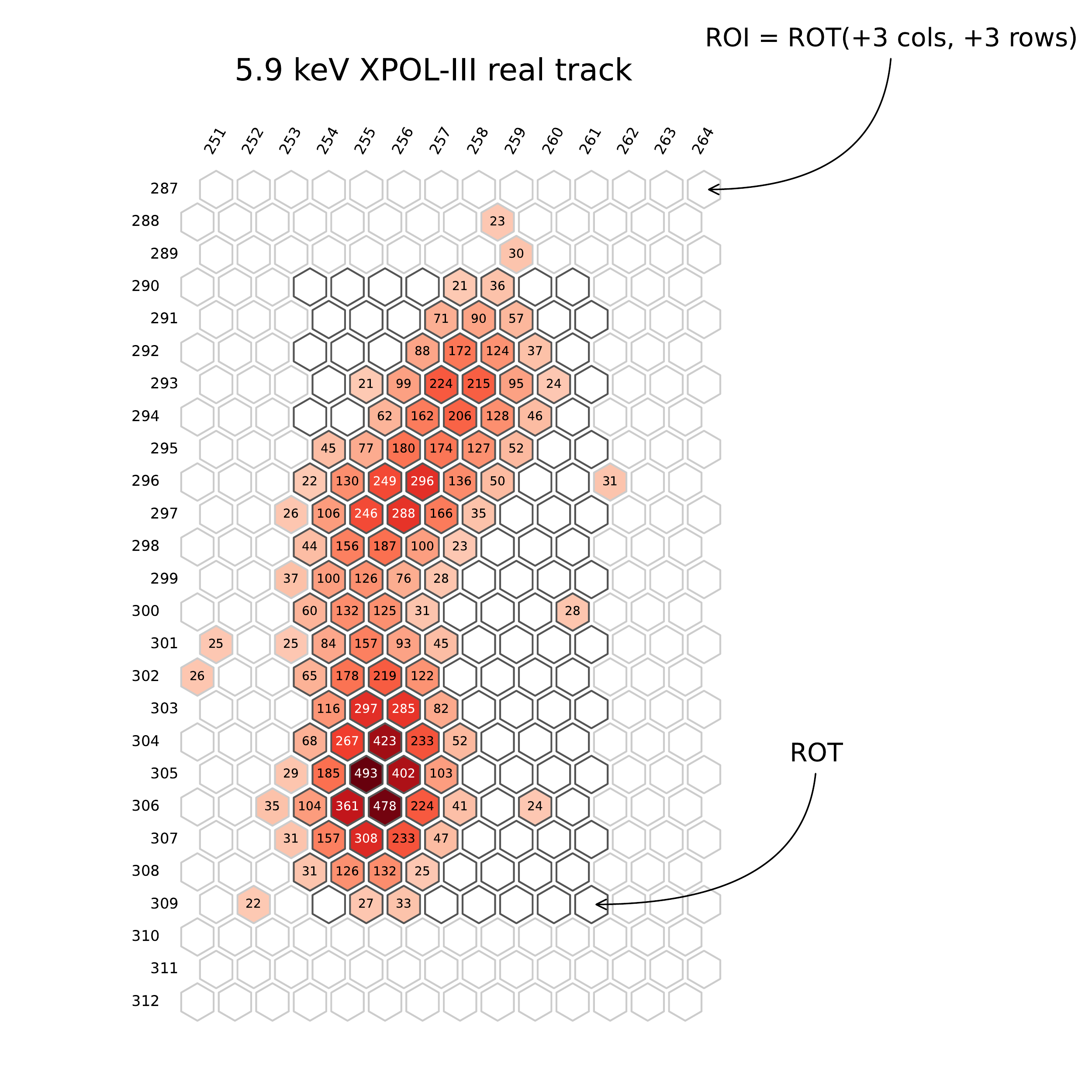}
    \caption{Display of a real \xpoliii{} track from a 5.9~keV radioactive $^{55}$Fe source
    with a padding of three pixels on all four sides of the ROT. The indices on the top and 
    left size of the ROI show the location on the readout matrix, while the text overlaid 
    on the pixels represent the corresponding PHA value. Even a superficial comparison with
    Figure~\ref{fig:xpoli_tracks} reveals how, given the much lower trigger threshold,
    the vast majority of the track is now contained in the ROT, and the padding of three
    pixels used here is in fact very generous even for a typical high energy track.}
    \label{fig:xpoliii_track}
\end{figure}

This section describes the initial electrical and functional tests that we performed 
to verify the basic functionality of the ASIC and gauge the optimal working point, in terms
of event padding and serial readout clock, to be used for measuring the relevant high-level
performance metrics that we shall present in section~\ref{sec:performance}.
For completeness, Figure~\ref{fig:xpoliii_track} shows a single event display of a real track
from a 5.9~keV radioactive $^{55}$Fe source, illustrating much of the improvements from  
the new trigger circuitry discussed in section~\ref{sec:architecture}.

\subsection{Electronics Noise}

\begin{figure}[!htb]
    \centering
    \includegraphics[width=\linewidth]{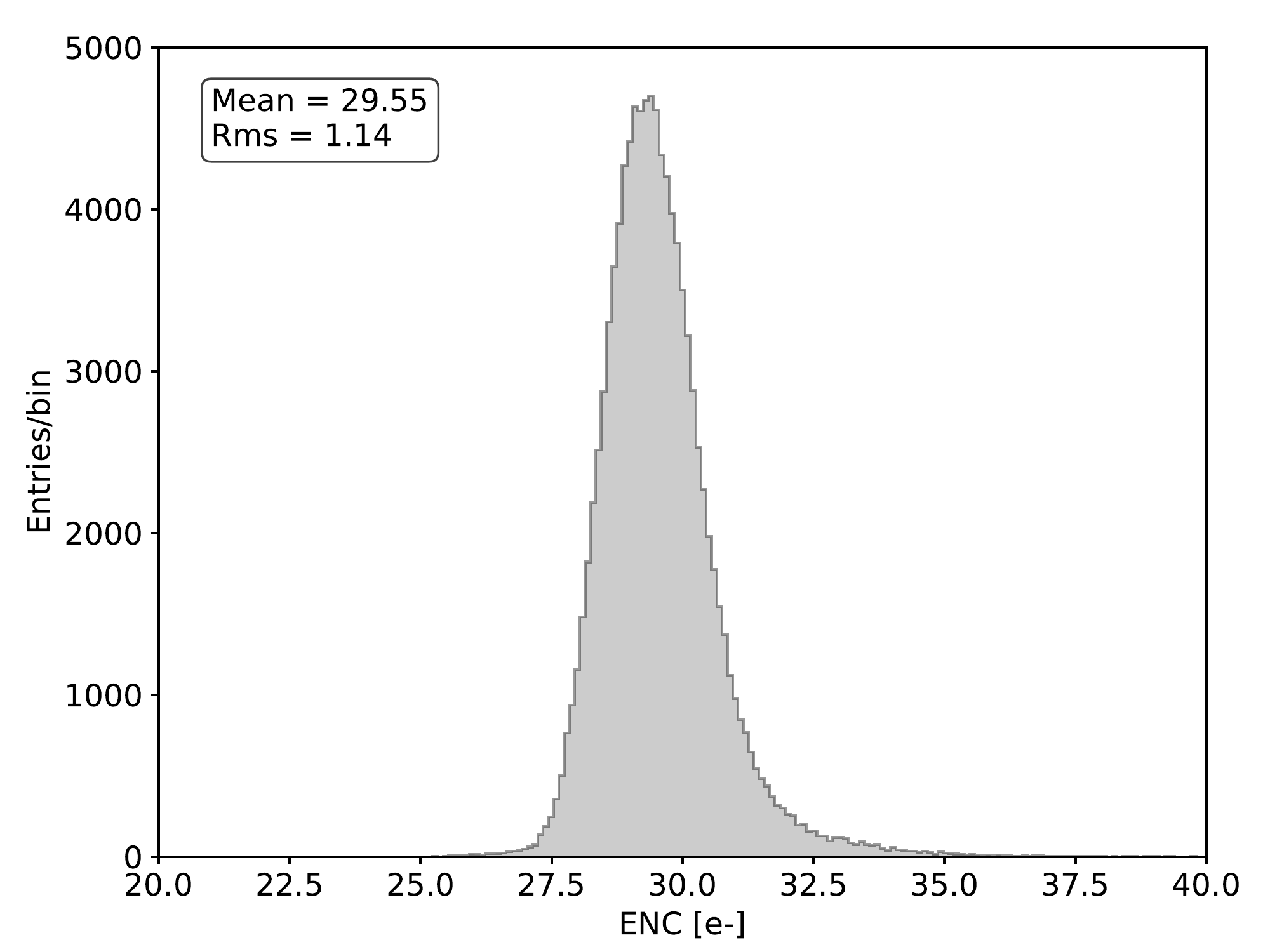}
    \caption{Noise distribution measured across the channels of a typical \xpoliii{} chip.
    The average noise is $\sim 30~e^{-}$~ENC.}
    \label{fig:noise_distribution}
\end{figure}

As already explained, and unlike the previous \xpol{} generations,
\xpoliii{} allows to set the coordinates of the ROI for the readout externally, and we 
extensively used this novel feature to measure the system noise in a number of readout chips.
More specifically, we use a series of partially overlapping, $21 \times 21$ regions of 
interest covering the entire active area of the chip and, for each ROI, we perform a 
number of readouts, with the same basic readout sequence used in nominal data-taking. 
For each pixel, the root mean square of the pedestal subtracted PHA values is a measure 
of the pre-amplifier noise%
\footnote{To be more precise, since we are operating with an online, event-by-event 
pedestal subtraction strategy as described in section~\ref{sec:architecture}, the raw rms 
values are a factor of $\sqrt{2}$ larger than the intrinsic noise of the amplifier.
This scale factor is correctly accounted for in Figure~\ref{fig:noise_distribution} as 
well as in the numbers quoted in the text.}

Figure~\ref{fig:noise_distribution} shows the noise distribution measured across the channels
of a typical \xpoliii{} chip. The average noise is about $30~e^{-}$~ENC. It should
be emphasized that this level of noise, translating into a track S/N ratio of $\sim 200$ for a
typical event, is largely irrelevant for our applications, and not a limiting factor in any
practical sense.

We note that the initial chips manufactured at the foundry and tested for this work have a small fraction of a \textperthousand{} dead channels; while this does not represent an issue for operating a GPD, a fine tuning of the manufacturing will be investigated with the foundry.

\subsection{ROT Padding}

We systematically investigated the effect of the ROT padding settings on the basic track 
properties by exploiting the hybrid \xpoliii{} operational mode described in 
section~\ref{sec:architecture} and acquiring X-ray data with different, pre-loaded padding 
values---from 2 to 5 pixels on all four sides of the ROT.
Since the main purpose of the padding is to guarantee the full containment of the electron
track, we resorted to using \emph{the fraction of events for which at least one of the pixels 
in the track lies on a border row or column}, which might indicate a possible track leakage,
as the relevant metric for our study. 

\begin{figure}[!hbt]
    \centering
    \includegraphics[width=0.9\linewidth]{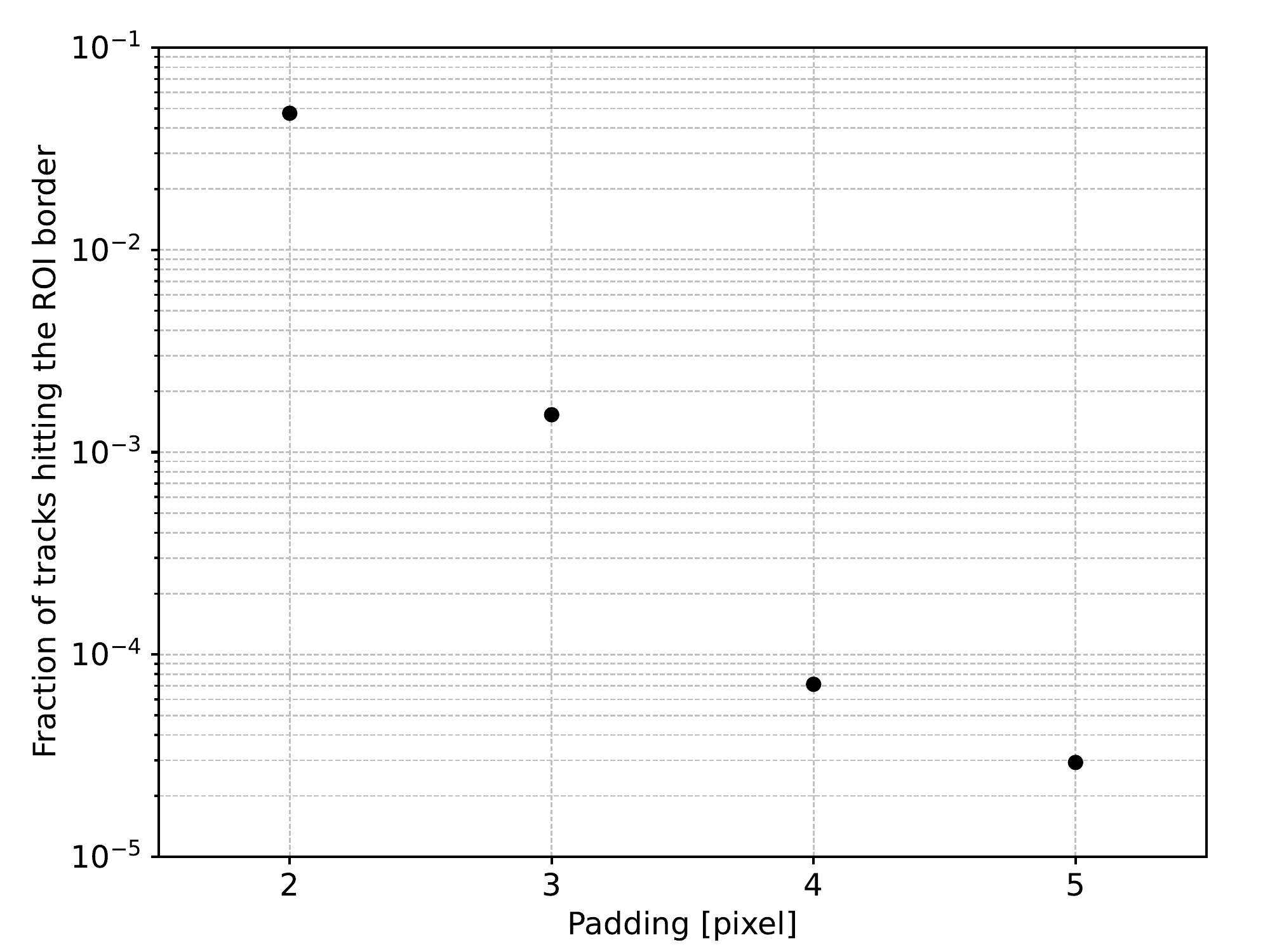}
    \caption{Fraction of events for which at least one of the pixels in the main track lies
    on the borders of the ROI, as s function of the ROT padding setting.
    The data points have been measured with a full irradiation of the detector with 5.9~keV
    photons from a $^{55}$Fe radioactive source, with the zero suppression threshold set to 
    $\sim 2\sigma$ of the noise.}
    \label{fig:margin_padding}
\end{figure}

As shown in Figure~\ref{fig:margin_padding} the fraction of tracks with border pixels
decreases monotonically as the padding increases, going from several \% at 
2~pixels to a few $10^{-5}$ at 5~pixels. We emphasize that a single pixel above threshold on
one of the borders of the ROI does not necessarily imply that the track has been actually 
truncated, nor that the polarimetric information encoded by the reconstructed track direction
has been compromised appreciably. 
We therefore conservatively take a padding 
value of 2~pixels as the smallest practically usable and we deem padding values larger than 
4 pixels as overly large, as the plot in Figure~\ref{fig:margin_padding} clearly indicates
that the curve starts flattening (cf. Figure~\ref{fig:roi_vs_energy}).

We note that more elaborated padding strategies (e.g., use different
values on different sides of the ROT) might conceivably provide an additional measurable
performance gain. The magnitude of the potential improvement can be gauged by noting that
reducing the padding by one unit on one side provides a relative reduction of the ROI size by
\begin{align}
    \frac{\delta n}{n} \approx \frac{1}{\sqrt{n}},
\end{align}
which is $\sim 7\%$ at a typical average ROI size of 200 pixels. 
Since investigating the effects of
such fine tuning is beyond the scope of this paper, we shall take a constant, comfortable padding of 
$3$ pixels as our nominal working point from now on, unless stated otherwise.

\subsection{Readout Time}

As explained in section~\ref{sec:readout}, the average readout time per event can be
parametrized as a linear function of the number of pixels in the ROI.
Since, for any given X-ray energy, events come in a large variety of different 
topologies, depending on the depth of the absorption point and the initial emission direction,
as well as the stochastic nature of the interaction processes-even photons from a monochromatic
source will generate in the detector ROIs of many different sizes and shapes. As a consequence,
illuminating the detector with a X-ray source provides a simple mean to study the dead time per
event as a function of the ROI size.

\begin{figure}[htb!]
    \centering
    \includegraphics[width=\linewidth]{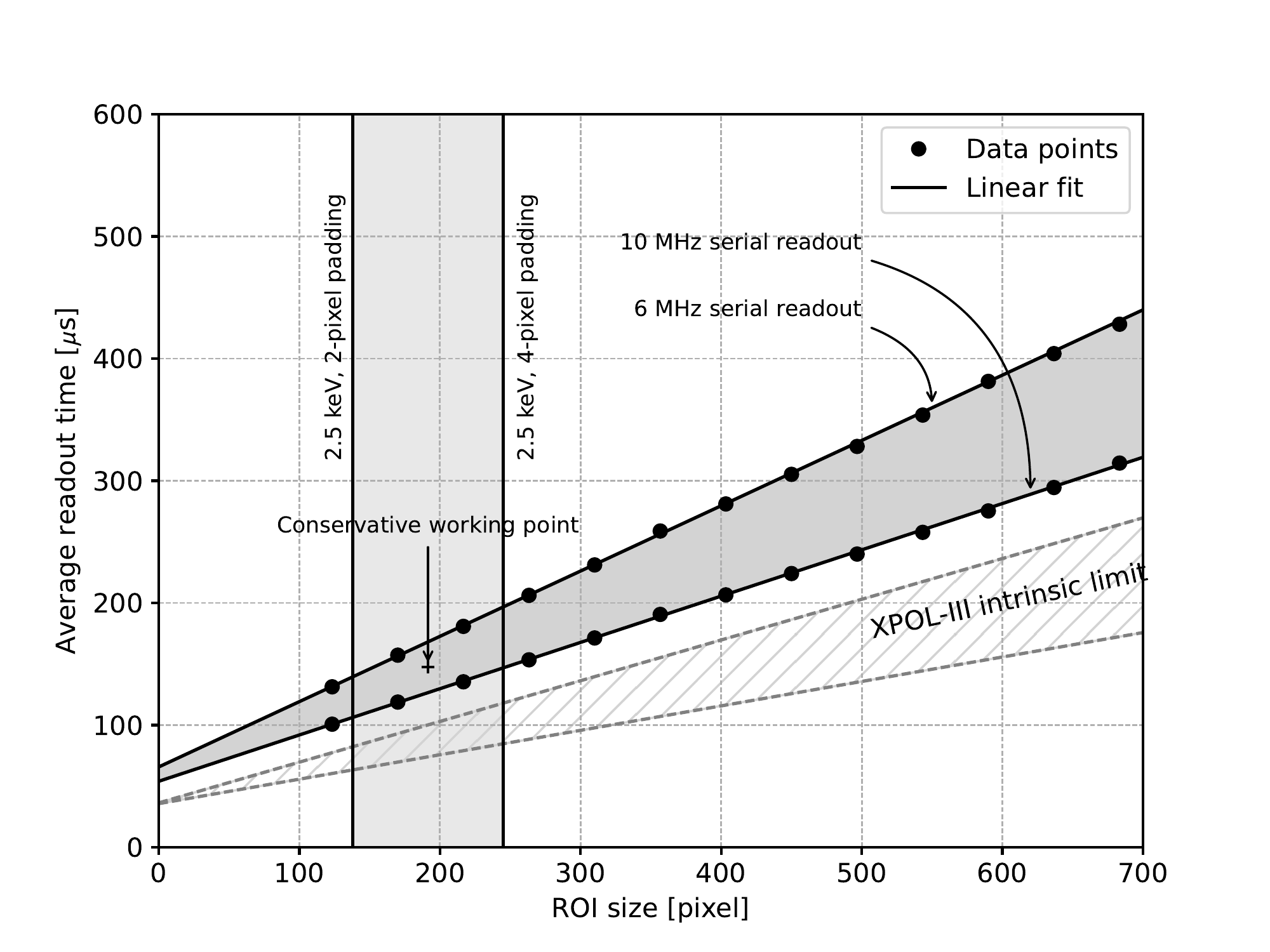}
    \caption{Readout time as a function of the ROI size, for two different values of the
    serial readout clock, obtained by illuminating the detector with 5.9~keV photons from
    a $^{55}$Fe radioactive source, with a padding of 3~pixels on all four sides of the ROT.
    The solid lines represent the best linear fits to the data points for two different values
    (6 and 10~MHz) of the serial readout clock. The vertical gray band indicates the average
    ROI size that we expect at the reference energy of 2.5~keV for a padding of 2 and 4 pixels.
    The hatched region between the two dashed lines indicate the intrinsic XPOL-III readout
    limit (at 6 and 10 MHz readout clock), assuming that the pedestal subtraction is fully
    parallelized and in the ideal case where the back-end electronics did not introduce any 
    additional delay to the readout sequence.}
    \label{fig:dead_time}
\end{figure}

Figure~\ref{fig:dead_time} shows the measured readout time as a function of the ROI size 
for two different values of the serial readout clock (6 and 10~MHz), and a constant padding 
of 3~pixels on all four sides of the ROT, with the detector uniformly illuminated with 5.9~keV
X-rays from a $^{55}$Fe radioactive source. The data are fitted with the straight line 
in Equation~\eqref{eq:readout_time}. At our reference energy of $\sim 2.5$~keV, the center of our fiducial 
region in the padding vs readout clock phase-space yields a tentative working point of $\sim 150~\mu$s,  at an average ROI size of $\sim 200$~pixels.
This is to be compared with $\sim 500$~pixels and $\sim 1$~ms for the \xpoli{} chip currently
operating on PolarLight and IXPE.

We note that, due to the current design of our back-end electronics, the measured readout 
time includes a contribution of $\sim 35~\mu$s for the pedestal subtraction in 
the FPGA, for a ROI of 200~pixels. 
For a high-throughput
application, this component could be conceivably parallelized at the cost of a slight increase
in the complexity of the system, yielding an additional $> 20$\% increase in readout speed without
side effects on any of the high-level performance metrics.

\section{High-Level Performance Figures}
\label{sec:performance}

In this section we briefly discuss the basic spectral and polarimetric response of the 
\xpoliii{} chip, embedded in the new GPD design. The reader should keep in mind that 
the readout ASIC is only a component of the mix, and the results are dependent on a number 
of other factors, including the detector assembly, the characteristic of the filling gas,
the amplification stage, as well as the track-reconstruction software.
For the sake of internal consistency, all the tests described in the following are done at the 
working point identified in section~\ref{sec:tests}, that is, with a constant padding of
3~pixels on the four sides of the ROT, and a serial readout clock of 7.5~MHz.

\subsection{Energy Resolution}

\begin{figure}[!hbt]
    \centering
    \includegraphics[width=0.9\linewidth]{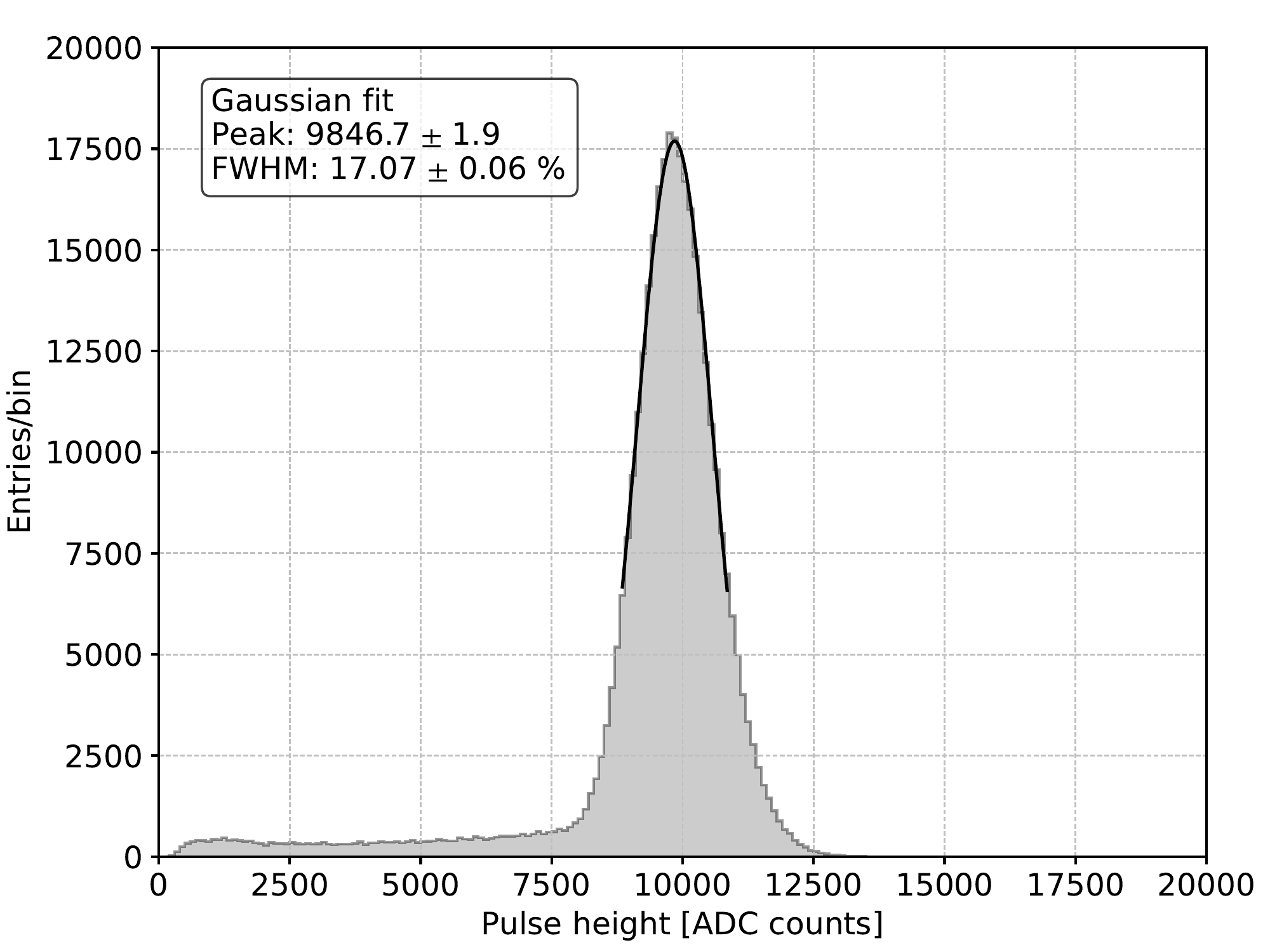}
    \caption{Spectrum of $^{55}$Fe line with a gaussian fit in nominal working conditions.
    The detector surface is uniformly illuminated and the large-scale gain dis-uniformity
    of the Gas Electron Multiplier corrected.}
    \label{fig:fe55peak}
\end{figure}

Figure~\ref{fig:fe55peak} shows the pulse-height distribution in pure DME at the nominal 
data taking settings for a flat field with 5.9~keV photons from a  $^{55}$Fe radioactive 
source. After correcting for spatial non-uniformities of the GEM, we achieve a FWHM 
around 17\%, in line with the corresponding figure for the IXPE flight 
detectors~\cite{BALDINI2021102628}.

% run 6247: 7.5MHz, padding 3, first ~0.5hr

We emphasize how even a superficial comparison with Figure~16 in~\cite{BALDINI2021102628} 
shows that the low-energy tail in the spectrum due to photon conversions in the passive materials
is reduced by a factor $\sim 2$, due to the elimination of the aluminum deposition on the inner 
face of the entrance window. While the suitability of a pure Be window for a long-term use in
space will be investigated and presented in a separate paper, we note that this factor is 
beneficial for both the spectral deconvolution and the polarization measurement.

\subsection{Azimuthal Response}

We tested the azimuthal response of the GPD assembly with both unpolarized photons from 
a 5.9~keV $^{55}$Fe radiactive source and polarized X-rays generated via Bragg diffraction at
45$^{\circ}$ on a graphite crystal. Our polarized setup produces three different monochromatic 
lines at 2.6, 5.2 and 7.8 keV, but the low- and high-energy ones are strongly suppressed due to
the absorption in air and the energy dependence of the GPD quantum efficiency, respectively.

%see plot_modulation.py for details
\begin{figure}[!hbt]
    \centering
    \includegraphics[width=0.9\linewidth]{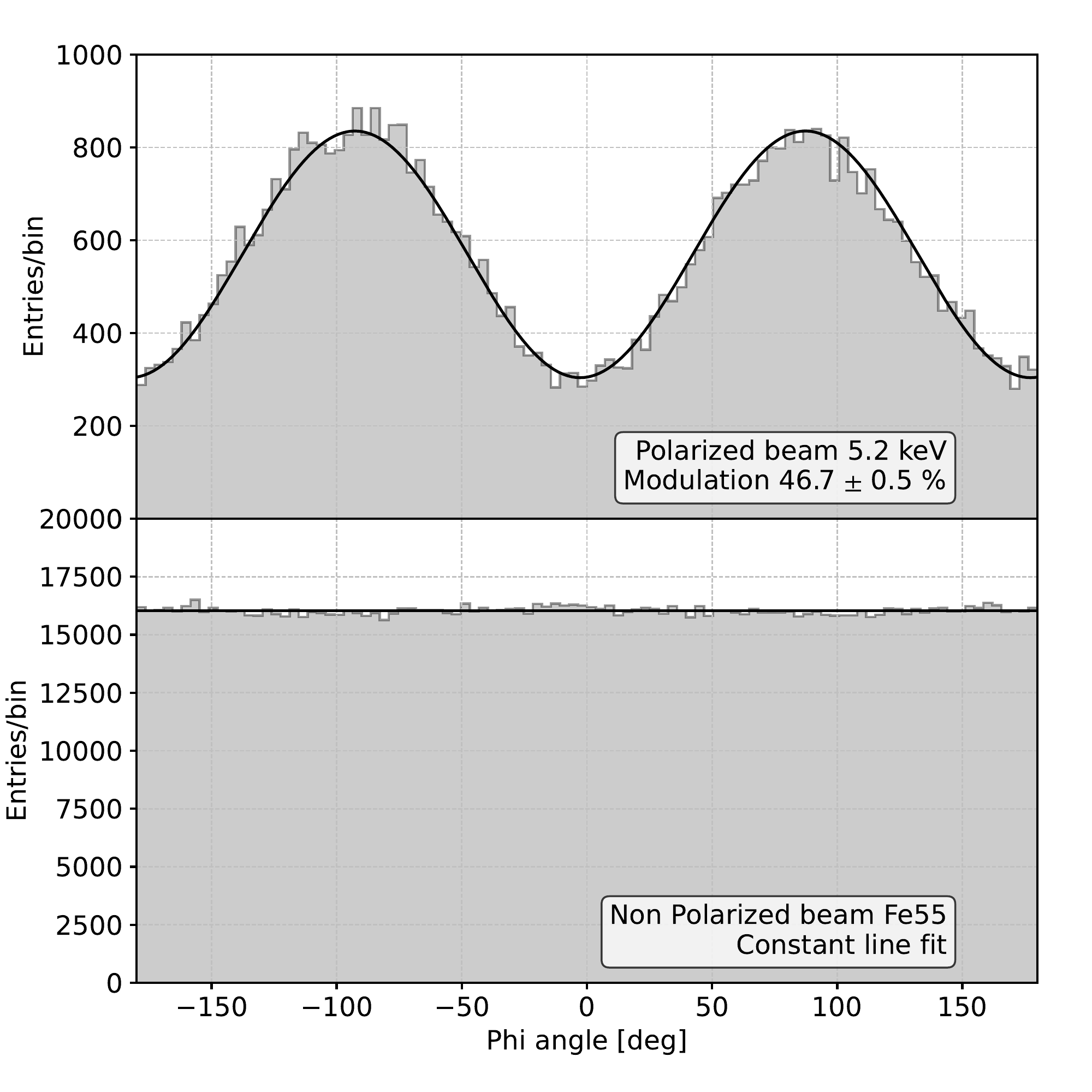}
    \caption{Measured modulation curve at 5.2 keV for a 100\% polarized X-ray source. 
    The modulation factor is 46.7\% without any selection on the events, in line with 
    previous measurements and with the prediction of our Monte Carlo simulation.}
    \label{fig:modulation_curves}
\end{figure}

Figure~\ref{fig:modulation_curves} shows two modulation curves, measured in our unpolarized
and polarized setups. For the latter, we selected events in the central 5.2~keV over a beam
spot with a $\sim 1$~mm radius close to the center of the detector, and performing two separate
acquisitions rotated by 90$^{\circ}$ in order to compensate for any possible residual systematic
effect. (Although, as shown in the bottom panel, such effects are largely subdominant.)
The measured modulation factor at 5.2~keV is $46.7 \pm 0.5$ \%, without any additional 
selection on the events other than that of the central line in the source spectrum.
The measured spurious modulation at 5.9~keV is smaller than 1\%, and, as explained in~\cite{BALDINI2021102628}, is 
to be ascribed to the multiplication stage.
This is in good agreement with both the figures for the IXPE focal-plane detectors (given the differences
in the GPD assembly and the X-ray source) and with the prediction from our Monte Carlo 
simulations, indicating that the new readout chip is fully preserving the intrinsic polarimetric
capabilities of the GPD. 

\section{Conclusions}

We have developed and tested a new custom readout chip for high-throughput X-ray
astronomical polarimetry. Compared to the ASIC currently operating on PolarLight and IXPE,
\xpoliii{} is able to generate significantly smaller, and yet fully contained track images
and to operate at a faster readout clock, with a streamlined readout sequence. 

When combined, these factors reduce the average deadtime per event to $\sim 150~\mu$s, 
or a factor of 7 smaller than the current state of the art. Considering that at least $\sim
35~\mu$s of the measured deadtime are an overhead introduced by the existing back-end
electronics, a speed-up of a full order of magnitude is clearly within reach, 
making  \xpoliii{} an ideal match for the upcoming generation of X-ray observatories.

This leap forward in readout speed is achieved with no measurable degradation in the polarimetric, spectral, imaging or timing capability of the detector, as demonstrated by the tests presented in this work. 

In addition, the reduction of at least a factor 2.5 of the ROI size is clearly very helpful in reducing the required bandwidth for both on board satellite operation as well for data transmission to the ground.

Finally, the possibility of operating at a trigger threshold as low as $\sim 150~e^{-}$ paves the way to new types of applications that were previously 
impossible, e.g., low-pressure configurations tailored to very low energies or, on the
other side of the energy band, operating in pure ionization mode with suitable gas mixtures.

We argue that this work constitutes a crucial step toward the deployment of a new generation
of GPD matching the needs of future X-ray observatories.
The development of suitable multiplications stages, free of systematic effects in the 
azimuthal response, is the other fundamental ingredient of the mix, and we shall report 
our progress in a separate paper.

\section{Acknowledgements}
This work was supported by the Italian Space Agency (ASI) through the agreements 
2018-11-HH.O, “ADAM - Advanced Detectors for X-ray Astronomy Mission" and 2017.13-H0, "Italian participation to the NASA IXPE mission"

\bibliographystyle{unsrt}
\bibliography{bibliography}

\begin{thebibliography}{1}

\bibitem{PolarLight}
Hua {Feng} and et~al.
\newblock {PolarLight: a CubeSat X-ray polarimeter based on the gas pixel
  detector}.
\newblock {\em Experimental Astronomy}, 47(1-2):225--243, April 2019.

\bibitem{IXPEJATIS}
Martin~C. Weisskopf and et~al.
\newblock The imaging x-ray polarimetry explorer (ixpe): Pre-launch, 2021.

\bibitem{BALDINI2021102628}
L.~Baldini and et~al.
\newblock Design, construction, and test of the gas pixel detectors for the
  ixpe mission.
\newblock {\em Astroparticle Physics}, 133:102628, 2021.

\bibitem{Costa2001662}
Costa E., Soffitta P., Bellazzini R., Brez A., Lumb N., and Spandre G.
\newblock An efficient photoelectric x-ray polarimeter for the study of black
  holes and neutron stars.
\newblock {\em Nature}, 411(6838):662 – 665, 2001.
\newblock Cited by: 274.

\bibitem{BELLAZZINI2004477}
R.~Bellazzini, F.~Angelini, L.~Baldini, F.~Bitti, A.~Brez, M.~Ceccanti,
  L.~Latronico, M.M. Massai, M.~Minuti, N.~Omodei, M.~Razzano, C.~Sgr\`o,
  G.~Spandre, E.~Costa, and P.~Soffitta.
\newblock Reading a gem with a vlsi pixel asic used as a direct charge
  collecting anode.
\newblock {\em Nuclear Instruments and Methods in Physics Research Section A:
  Accelerators, Spectrometers, Detectors and Associated Equipment}, 535(1):477
  -- 484, 2004.
\newblock Proceedings of the 10th International Vienna Conference on
  Instrumentation.

\bibitem{BELLAZZINI2006552}
R.~Bellazzini, G.~Spandre, M.~Minuti, L.~Baldini, A.~Brez, F.~Cavalca,
  L.~Latronico, N.~Omodei, M.M. Massai, C.~Sgr\`o, E.~Costa, P.~Soffitta,
  F.~Krummenacher, and R.~{de Oliveira}.
\newblock Direct reading of charge multipliers with a self-triggering cmos
  analog chip with 105k pixels at 50$\mu$m pitch.
\newblock {\em Nuclear Instruments and Methods in Physics Research Section A:
  Accelerators, Spectrometers, Detectors and Associated Equipment}, 566(2):552
  -- 562, 2006.

\bibitem{Ramsey}
Robert~A. Austin and Brian~D. Ramsey.
\newblock {Optical imaging chamber for x-ray astronomy}.
\newblock {\em Optical Engineering}, 32(8):1990 -- 1994, 1993.

\bibitem{eXTP}
S.~N. Zhang and et~al.
\newblock {eXTP: Enhanced X-ray Timing and Polarization mission}.
\newblock In Jan-Willem~A. den Herder, Tadayuki Takahashi, and Marshall Bautz,
  editors, {\em Space Telescopes and Instrumentation 2016: Ultraviolet to Gamma
  Ray}, volume 9905, pages 505 -- 520. International Society for Optics and
  Photonics, SPIE, 2017.

\bibitem{BarbaneraTNS}
M.~Barbanera, S.~Citraro, C.~Magazz\`u, A.~Manfreda, M.~Minuti, H.~Nasimi, and
  C.~Sgr\`o.
\newblock Initial tests and characterization of the readout electronics for the
  ixpe mission.
\newblock {\em IEEE Transactions on Nuclear Science}, 68(5):1144--1151, 2021.

\end{thebibliography}

\end{document}